\documentclass{article}
\usepackage{amssymb}
\usepackage{amsmath}

\input amssym.def
\input amssym

\newtheorem{theorem}{Theorem}
\newtheorem{lemma}{Lemma}

\newcommand{\C}{\mathbb C} 
\newcommand{\R}{\mathbb R} 
\newcommand{\N}{\mathbb N} 

\newcommand{\mB}{\mathcal{B}} 

\newcommand{\mT}{\mathcal{T}}

\newcommand{\hi}{\mathcal{H}} 
\newcommand{\tsh}{\mathcal{T}_s(\hi)} 
\newcommand{\bsh}{\mathcal{B}_s(\hi)} 
\newcommand{\csh}{\mathcal{C}_s(\hi)} 
\newcommand{\sh}{\mathcal{S}(\hi)} 
\newcommand{\esh}{\partial_e\sh} 
\newcommand{\ph}{\mathcal{P}(\hi)} 

\newcommand{\tr}[1]{\mathrm{tr} \, {#1}} 
\newcommand{\no}[1]{\left\|{#1}\right\|} 
\newcommand{\kb}[2]{|#1\,\rangle\langle\,#2|} 
\newcommand{\ip}[2]{\langle {#1}|{#2}\rangle} 

\newcommand{\fii}{\varphi}

\newenvironment{proof}[1][
Proof]{\textbf{#1.} }{\ \rule{0.5em}{0.5em}}
\newcommand{\qed}{\ $\square$}

\begin{document}

\title{Topologies and Measurable Structures on the Projective
Hilbert Space\thanks{This work was supported by the Perimeter Institute
for Theoretical Physics, Waterloo, Ontario, Canada.}}

\author{Werner Stulpe\thanks{Electronic address: stulpe@fh-aachen.de}\\
{\small Aachen University of Applied Sciences, J\"ulich Campus,
D-52428, Germany}}

\date{}
\maketitle

\begin{abstract}
\noindent A systematic review of the various topologies that can be defined
on the projective Hilbert space $ \ph $, i.e., on the set of the pure
quantum states, is presented. It is shown that $ \ph $ carries a natural
topology as well as a natural measurable structure.\\
\mbox{}\\
Key words: Projective Hilbert space, pure quantum states, topology and
Borel structure.
\end{abstract}

\section{Introduction}

Based on ideas of Misra \cite{mis74}, it was essentially the late
S. Bugajski who recognized that the statistical (probabilistic) framework
of quantum mechanics can be understood as a reduced classical
probability theory on the projective Hilbert space $ \ph $
\cite{bug91;93a-d,bel95a;b,hol82,stu01,bus04}. In a forthcoming paper
\cite{bus07} a suggestive definition of a classical extension of
quantum mechanics is given and it is proved that every such classical
extension is essentially equivalent to the Misra-Bugajski scheme. Moreover,
it is known that $ \ph $, considered as a real differentiable manifold,
carries a Riemannian as well as a symplectic structure,
the latter enabling one to reformulate quantum dynamics
on the (in general infinite-dimensional) phase space $ \ph $
\cite{gue77,kib79,cir84,cir90a;b,bro01,bje05}. Thus, quantum mechanics
can be considered as some reduced classical statistical mechanics.

In order to take the projective Hilbert space $ \ph $ as a sample space
for classical probability theory, $ \ph $ must be equipped
with a measurable structure. To make $ \ph $ a differentiable manifold,
it should be equipped with a topology first. The elements of $ \ph $ can
be interpreted as equivalence classes of vectors $ \fii \in \hi $,
$ \fii \neq 0 $, as equivalence classes of unit vectors, as the
one-dimensional subspaces of $ \hi $, or as the one-dimensional
orthogonal projections acting in $ \hi $. So, on the one hand,
two qoutient topologies can be defined on $ \ph $ whereas,
on the other hand, all the different operator topologies induce
topologies on $ \ph $. In Section 2 we undertake a systematic review
and comparison, already sketched out in \cite{bug94}, of the various
topologies on $ \ph $ and show that all the topologies are the same,
thus giving a natural topology $ \mT $ on $ \ph $. Furthermore,
in Section 3 we present a simple proof that the Borel structure
generated by the topology $ \mT $ coincides with the measurable structure
generated by the transition probability functions of the pure
quantum states; our proof simplifies a proof of Misra from 1974 for
a corresponding statement \cite{mis74}.

\section{The Topology of the Projective Hilbert Space}\label{sec:top}

Let $ \hi \neq \{ 0 \} $ be a nontrivial separable complex
Hilbert space. Call two vectors of $ \hi^* := \hi \setminus \{0\} $
equivalent if they differ by a complex factor, and define the
{\it projective Hilbert space $ \ph $} to be the set of the corresponding
equivalence classes which are often called {\it rays}. Instead of
$ \hi^* $ one can consider only the unit sphere of $ \hi $,
$ S := \{\fii \in \hi \, | \, \no{\fii} = 1 \} $. Then two unit vectors
are called equivalent if they differ by a phase factor, and the set of
the corresponding equivalence classes, i.e., the set of the {\it
unit rays}, is denoted by $ S/S^1 $ (in this context, $ S^1 $ is
understood as the set of all phase factors, i.e., as the set of all
complex numbers of modulus $1$). Clearly, $ S/S^1 $ can be
identified with the projective Hilbert space $ \ph $. Furthermore,
we can consider the elements of $ \ph $ also as the one-dimensional
subspaces of $ \hi $ or, equivalently, as the one-dimensional
orthogonal projections $ P = P_{\fii} = \kb{\fii}{\fii} $, $
\no{\fii} = 1 $.

The set $ \hi^* $ and the unit sphere $S$ carry the topologies induced by
the metric topology of $ \hi $. Using the canonical projections
$ \mu \! : \hi^* \to \ph $, $ \mu(\fii) := [\fii] $, and
$ \nu \! : S \to S/S^1 $, $ \nu(\chi) := [\chi]_S $, where
$ [\fii] $ is a ray and $ [\chi]_S $ a unit ray, we can equip
the quotient sets $ \ph $ and $ S/S^1 $ with their quotient topologies
$ \mT_{\mu} $ and $ \mT_{\nu} $. Considering $ \mT_{\nu} $, a set
$ O \subseteq S/S^1 $ is called open if $ \nu^{-1}(O) $ is open.

\begin{theorem}\label{thm:ss1}
The set $ S/S^1 $, equipped with the quotient topology $ \mT_{\nu} $,
is a second-countable Hausdorff space, and $ \nu $ is an open continuous
mapping.
\end{theorem}
\proof{
By definition of $ \mT_{\nu} $, $ \nu $ is continuous. To show that
$ \nu $ is open, let $U$ be an open set of $S$. From
\[
\nu^{-1}(\nu(U)) = \nu^{-1}(\{ [\chi]_S \, | \, \chi \in U \})
                 = \bigcup_{\lambda \in S^1} \lambda U ,
\]
$ S^1 = \{ \lambda \in \C \, | \, |\lambda| = 1 \} $, it follows that
$ \nu^{-1}(\nu(U)) \subseteq S $ is open. So $ \nu(U) \subseteq S/S^1 $
is open; hence, $ \nu $ is open.

Next consider two different unit rays $ [\fii]_S $ and $ [\psi]_S $ where
$ \fii,\psi \in S $ and $ |\ip{\fii}{\psi}| = 1 - \varepsilon $,
$ 0 < \varepsilon \leq 1 $. Since the mapping
$ \chi \mapsto |\ip{\fii}{\chi}| $, $ \chi \in S $, is continuous,
the sets
\begin{equation}\label{u1}
U_1 := \left\{ \chi \in S \left| \,
              |\ip{\fii}{\chi}| > 1 - \tfrac{\varepsilon}{2} \right.
                                                             \right\}
\end{equation}
and
\begin{equation}\label{u2}
U_2 := \left\{ \chi \in S \left| \,
              |\ip{\fii}{\chi}| < 1 - \tfrac{\varepsilon}{2} \right.
                                                             \right\}
\end{equation}
are open neighborhoods of $ \fii $ and $ \psi $, respectively. Consequently,
the sets $ O_1 := \nu(U_1) $ and $ O_2 := \nu(U_2) $ are open
neighborhoods of $ [\fii]_S $ and $ [\psi]_S $, respectively. Assume
$ O_1 \cap O_2 \neq \emptyset $. Let $ [\xi]_S \in O_1 \cap O_2 $, then
$ [\xi]_S = \nu(\chi_1) = \nu(\chi_2) $ where $ \chi_1 \in U_1 $ and
$ \chi_2 \in U_2 $. It follows that $ \chi_1 $ and $ \chi_2 $ are equivalent,
so $ |\ip{\fii}{\chi_1}| = |\ip{\fii}{\chi_2}| $, in contradiction to
$ \chi_1 \in U_1 $ and $ \chi_2 \in U_2 $. Hence, $ O_1 $ and $ O_2 $
are disjoint, and $ \mT_{\nu} $ is separating.

Finally, let $ \mB = \{ U_n \, | \, n \in \N \} $ be a countable base of
the topology of $S$ and define the open sets $ O_n := \nu(U_n) $. We show
that $ \{ O_n \, | \, n \in \N \} $ is a base of $ \mT_{\nu} $. For
$ O \in \mT_{\nu} $, we have that $ \nu^{-1}(O) $ is an open set of
$S$ and consequently $ \nu^{-1}(O) = \bigcup_{n \in M} U_n $ where
$ U_n \in \mB $ and $ M \subseteq \N $. Since $ \nu $ is surjective,
it follows that
\[
O = \nu(\nu^{-1}(O)) = \nu \left( \bigcup_{n \in M} U_n \right)
                     = \bigcup_{n \in M} \nu(U_n)
                     = \bigcup_{n \in M} O_n .
\]
Hence, $ \{ O_n \, | \, n \in \N \} $ is a countable base of
$ \mT_{\nu} $. \qed}

Analogously, it can be proved that the topology $ \mT_{\mu} $ on $ \ph $
is separating and second-countable and that the canonical projection $ \mu $
is open (and continuous by the definition of $ \mT_{\mu} $). Moreover,
one can show that the natural bijection $ \beta \! : \ph \to S/S^1 $,
$ \beta([\fii]) := \left[ \frac{\fii}{\no{\fii}} \right]_S $,
$ \beta^{-1}([\chi]_S) = [\chi] $, is a homeomorphism. Thus, identifying
$ \ph $ and $ S/S^1 $ by $ \beta $, the topologies $ \mT_{\mu} $ and
$ \mT_{\nu} $ are the same.

We denote the real vector space of the self-adjoint trace-class operators by
$ \tsh $ and the real vector space of all bounded self-adjoint operators by
$ \bsh $; endowed with the trace norm and the usual operator norm,
respectively, these spaces are Banach spaces. As is well known,
$ \bsh $ can be considered as the dual space $ (\tsh)' $ where the duality
is given by the trace functional. Let $ \sh $ be the convex set of
all positive trace-class operators of trace $1$; the operators of
$ \sh $ are the density operators and describe the quantum states. We
recall that the extreme points of the convex set $ \sh $, i.e.,
the pure quantum states, are the one-dimensional orthogonal projections
$ P = P_{\fii} $, $ \no{\fii} = 1 $. We denote the set of these extreme
points, i.e., the extreme boundary, by $ \esh $.

The above definition of $ \ph $ and $ S/S^1 $ as well as of their
quotient topologies is related to a geometrical point of view. From
an operator-theoretical point of view, it is more obvious to identify
$ \ph $ with $ \esh $ and to restrict one of the various operator topologies
to $ \esh $. A further definition of a topology on $ \esh $ is suggested by
the interpretation of the one-dimensional projections $ P \in \esh $ as
the pure quantum states and by the requirement that
the transition probabilities between two pure states are
continuous functions. Next we consider, taking account of
$ \esh \subseteq \sh \subset \tsh \subseteq \bsh $, the metric topologies on
$ \esh $ induced by the trace-norm topology of $ \tsh $, resp.,
by the norm toplogy of $ \bsh$. After that we introduce the weak topology on
$ \esh $ defined by the transition-probability functions as well as
the restrictions of several weak operator topologies to $ \esh $. Finally,
we shall prove the surprising result that all the many toplogies on
$ \ph \cong S/S^1 \cong \esh $ are equivalent.

\begin{theorem}\label{thm:esh-dist}
Let $ P_{\fii} = \kb{\fii}{\fii} \in \esh $ and
$ P_{\psi} = \kb{\psi}{\psi} \in \esh $ where
$ \no{\fii} = \no{\psi} = 1 $. Then
\begin{enumerate}
\item[(a)]
\[ \rho_n(P_{\fii},P_{\psi}) := \no{P_{\fii} - P_{\psi}}
                              = \sqrt{1 - |\ip{\fii}{\psi}|^2}
                              = \sqrt{1 - \tr{P_{\fii} P_{\psi}}}
\]
where the norm $ \no{\cdot} $ is the usual operator norm
\item[(b)]
\[ \rho_{\mathrm{tr}}(P_{\fii},P_{\psi})
               :=  \no{P_{\fii} - P_{\psi}}_{\mathrm{tr}}
                = 2\no{P_{\fii} - P_{\psi}},
\]
in particular, the metrics $ \rho_n $ and $ \rho_{\mathrm{tr}} $ on
$ \esh $ induced by the operator norm $ \no{\cdot} $ and the trace
norm $ \no{\cdot}_{\mathrm{tr}} $ are equivalent
\item[(c)]
\[
\no{P_{\fii} - P_{\psi}} \leq \no{\fii - \psi},
\]
in particular, the mapping $ \fii \mapsto P_{\fii} $ from $S$ into
$ \esh $ is continuous, $ \esh $ being equipped with $ \rho_n $ or
$ \rho_{\mathrm{tr}} $.
\end{enumerate}
\end{theorem}
\proof{
To prove (a) and (b), assume $ P_{\fii} \neq P_{\psi} $, otherwise
the statements are trivial. Then the range of $ P_{\fii} - P_{\psi} $
is a two-dimensional subspace of $ \hi $ and is spanned by the two linearly
independent unit vectors $ \fii $ and $ \psi $. Since eigenvectors of
$P_{\fii} - P_{\psi} $ belonging to eigenvalues $ \lambda \neq 0 $
must lie in the range of $ P_{\fii} - P_{\psi} $, they can be written as
$ \chi = \alpha \fii + \beta \psi $. Therefore, the eigenvalue problem
$ (P_{\fii} - P_{\psi}) \chi = \lambda \chi $, $ \chi \neq 0 $,
is equivalent to the two linear equations
\begin{eqnarray*}
    (1- \lambda) \alpha + \ip{\fii}{\psi} \beta & = & 0   \\
-\ip{\psi}{\fii} \alpha -   (1 + \lambda) \beta & = & 0
\end{eqnarray*}
where $ \alpha \neq 0 $ or $ \beta \neq 0 $. It follows that
$ \lambda = \pm \sqrt{1 - |\ip{\fii}{\psi}|^2} =: \lambda_{1,2} $. Hence,
$ P_{\fii} - P_{\psi} $ has the eigenvalues
$ \lambda_1 $, $0$, and $ \lambda_2 $. Now, from
$ \no{P_{\fii} - P_{\psi}} = \max \{ |\lambda_1|,|\lambda_2| \} $ and
$ \no{P_{\fii} - P_{\psi}}_{\mathrm{tr}} = |\lambda_1| + |\lambda_2| $,
we obtain the statements (a) and (b).---From
\begin{eqnarray*}
\no{P_{\fii} - P_{\psi}}^2
&  =   & 1 - |\ip{\fii}{\psi}|^2
   =     \no{\fii - \ip{\psi}{\fii} \psi}^2
   =     \no{(I - P_{\psi}) \fii}^2                                  \\
& \leq & \no{(I - P_{\psi}) \fii}^2 + \no{\psi - P_{\psi} \fii}^2    \\
&  =   & \no{(I - P_{\psi}) \fii - (\psi - P_{\psi} \fii)}^2         \\
&  =   & \no{\fii - \psi}^2
\end{eqnarray*}
we conclude statement (c). \qed}

According to statement (b) of Theorem \ref{thm:esh-dist}, the metrics $ \rho_n $ and
$ \rho_{\mathrm{tr}} $ give rise to the same topology
$ {\mT}_n = {\mT}_{\mathrm{tr}} $ as well as to the same uniform structures.

\begin{theorem}\label{thm:esh-sepc}
Equipped with either of the two metrics $ \rho_n $ and
$ \rho_{\mathrm{tr}} $, $ \esh $ is separable and complete.
\end{theorem}
\proof{
As a metric subspace of the separable Hilbert space $ \hi $, the unit sphere
$S$ is separable. Therefore, by statement (c) of Theorem \ref{thm:esh-dist}, the metric space
$ (\esh,\rho_n) $ is separable and so is $ (\esh,\rho_{\mathrm{tr}}) $
(the latter, moreover, implies the trace-norm separability of
$ \tsh $). Now let $ \{ P_n \}_{n \in \N} $ be a Cauchy sequence in
$ (\esh,\rho_{\mathrm{tr}}) $. Then there exists an operator
$ A \in \tsh $ such that $ \no{P_n - A}_{\mathrm{tr}} \to 0 $ as well as
$\no{P_n - A} \to 0 $ as $ n \to \infty $ (remember that, on $ \tsh $,
$ \no{\cdot}_{\mathrm{tr}} $ is stronger than $ \no{\cdot} $). From
\begin{eqnarray*}
\no{P_n - A^2}    =     \no{A^2 - P_n^2}
               & \leq & \no{A^2 - AP_n} + \no{AP_n - P_n^2}    \\
               & \leq & \no{A} \no{A - P_n} + \no{A - P_n}     \\
               & \to  &  0
\end{eqnarray*}
as $ n \to \infty $ we obtain $ A = \lim_{n \to \infty} P_n = A^2 $;
moreover,
\[
\tr{A} = \tr{AI} = \lim_{n \to \infty} \tr{P_nI} = 1.
\]
Hence, $A$ is a one-dimensional orthogonal projection, i.e.,
$ A \in \esh $. \qed}

Next we equip $ \esh $ with the topology $ \mT_0 $ generated by
the functions
\begin{equation}\label{p}
P \mapsto h_Q(P) := \tr{PQ} = |\ip{\fii}{\psi}|^2
\end{equation}
where $ P = \kb{\psi}{\psi} \in \esh $, $ Q = \kb{\fii}{\fii} \in \esh $,
and $ \no{\psi} = \no{\fii} = 1 $. That is, $ \mT_0 $ is the coarsest
topology on $ \esh $ such that all the real-valued functions $ h_Q $
are continuous. Note that $ \tr{PQ} = |\ip{\fii}{\psi}|^2 $ can be
interpreted as the transition probability between the two pure states
$P$ and $Q$.

\begin{lemma}\label{lem:esh}
The set $ \esh $, equipped with the topology $ \mT_0 $, is
a second-countable Hausdorff space. A countable base of $ \mT_0 $
is given by the finite intersections of the open sets
\begin{equation}\label{uklm}
\begin{array}{crl}
U_{klm} & := & h_{Q_k}^{-1} \left( \,
                            \left] q_l - \frac{1}{m},q_l + \frac{1}{m}
                                         \right[ \, \right)   \vspace{2mm}\\
        &  = & \left\{ P \in \esh \left| \,
               \left| \tr{PQ_k} - q_l \right| < \frac{1}{m}
                                                \right. \right\}
\end{array}
\end{equation}
where $ \{ Q_k \}_{k \in \N} $ is a sequence of one-dimensional orthogonal
projections being $ \rho_n $-dense in $ \esh $, $ \{ q_l \}_{l \in \N} $
is a sequence of numbers being dense in $ [0,1] \subseteq \R $, and
$ m \in \N $.
\end{lemma}
\proof{
Let $ P_1 $ and $ P_2 $ be any two different one-dimensional
projections. Choosing $ Q = P_1 $ in (\ref{p}), we obtain
$ h_{P_1}(P_1) = 1 \neq h_{P_1}(P_2) = 1 - \varepsilon $,
$ 0 < \varepsilon \leq 1 $. The sets
\[
U_1 := \left\{ P \in \esh \left| \,
                h_{P_1}(P) > 1 - \tfrac{\varepsilon}{2} \right. \right\}
\]
and
\[
U_2 := \left\{ P \in \esh \left| \,
                h_{P_1}(P) < 1 - \tfrac{\varepsilon}{2} \right. \right\}
\]
(cf.\ Eqs.~(\ref{u1}) and (\ref{u2})) are disjoint open neighborhoods of $ P_1 $ and
$ P_2 $, respectively. So $ \mT_0 $ is separating.

For an open set $ O \subseteq \R $, $ h_Q^{-1}(O) $ is $ \mT_0 $-open. We
next prove that
\begin{equation}\label{u}
U := h_Q^{-1}(O) = \bigcup_{U_{klm} \subseteq U} U_{klm}
\end{equation}
with $ U_{klm} $ according to (\ref{uklm}). Let $ P \in U $. Then
there exists an $ \varepsilon > 0 $ such that the interval
$ ]h_Q(P) - \varepsilon,h_Q(P) + \varepsilon[ $ is contained in $O$. Choose
$ m_0 \in \N $ such that $ \frac{1}{m_0} < \frac{\varepsilon}{2} $,
and choose a member $ q_{l_0} $ of the sequence $ \{ q_l \}_{l \in \N} $
and a member $ Q_{k_0} $ of $ \{ Q_k \}_{k \in \N} $ such that
$ |\tr{PQ} - q_{l_0}| < \frac{1}{2m_0} $ and
$ \no{Q_{k_0} - Q} < \frac{1}{2m_0} $. It follows that
\begin{eqnarray*}
|\tr{PQ_{k_0}} - q_{l_0}|
& \leq & |\tr{PQ_{k_0}} - \tr{PQ}| + |\tr{PQ} - q_{l_0}|    \\
& \leq &  \no{Q_{k_0} - Q} + |\tr{PQ} - q_{l_0}|            \\
&  <   &  \tfrac{1}{m_0}
\end{eqnarray*}
which, by (\ref{uklm}), means that $ P \in U_{k_0l_0m_0} $. We further have
to show that $ U_{k_0l_0m_0} \subseteq U $. To that end, let
$ \widetilde{P} \in U_{k_0l_0m_0} $. Then, from
\[
   \bigl| \tr{\widetilde{P}Q} - \tr{PQ} \bigr| \leq
   \bigl| \tr{\widetilde{P}Q} - \tr{\widetilde{P}Q_{k_0}} \bigr|
 + \bigl| \tr{\widetilde{P}Q_{k_0}} - q_{l_0} \bigr| + |q_{l_0} - \tr{PQ}|
\]
where the first term on the right-hand side is again smaller than
$ \no{Q - Q_{k_0}} $ and, by (\ref{uklm}), the second term is smaller than
$ \frac{1}{m_0} $, it follows that
\[
\bigl| h_Q(\widetilde{P}) - h_Q(P) \bigr|
 =   \bigl| \tr{\widetilde{P}Q} - \tr{PQ} \bigr|
\leq \tfrac{1}{2m_0} + \tfrac{1}{m_0} + \tfrac{1}{2m_0}
 = \tfrac{2}{m_0}
 < \varepsilon.
\]
This implies that $ h_Q(\widetilde{P}) \in
{]h_Q(P) - \varepsilon,h_Q(P) + \varepsilon[} \subseteq O $, i.e.,
$ \widetilde{P} \in h_Q^{-1}(O) = U $. Hence,
$ U_{k_0l_0m_0} \subseteq U $.

Summarizing, we have shown that, for $ P \in U $,
$ P \in U_{k_0l_0m_0} \subseteq U $. Hence,
$ U \subseteq \bigcup_{U_{klm} \subseteq U} U_{klm} \subseteq U $,
and assertion (\ref{u}) has been proved. The finite intersections of sets
of the form $ U = h_Q^{-1}(O) $ constitute a basis of the topology
$ \mT_0 $. Since every set $ U = h_Q^{-1}(O) $ is the union of sets
$ U_{klm} $, the intersections of finitely many sets
$ U = h_Q^{-1}(O) $ is the union of finite intersections of sets
$ U_{klm} $. Thus, the finite intersections of the sets
$ U_{klm} $ constitute a countable base of $ \mT_0 $. \qed}

Later we shall see that the topological space $ (\esh,\mT_0) $ is
homeomorphic to $ (\esh,\mT_n) $ as well as to $ (S/S^1,\mT_{\nu})
$. So it is also clear by Theorem \ref{thm:esh-sepc} or Theorem
\ref{thm:ss1} that $ (\esh,\mT_0) $ is a second-countable Hausdorff
space. The reason for stating Lemma \ref{lem:esh} is that later we
shall make explicit use of the particular countable base given
there.

The weak operator topology on the space $ \bsh $ of the bounded
self-adjoint operators on $ \hi $ is the coarsest topology such that
the linear functionals
\[
A \mapsto \ip{\fii}{A\psi}
\]
where $ A \in \bsh $ and $ \fii,\psi \in \hi $, are continuous. It is
sufficient to consider only the functionals
\begin{equation}\label{afun}
A \mapsto \ip{\fii}{A\fii}
\end{equation}
where $ \fii \in \hi $ and $ \no{\fii} = 1 $. The topology $ \mT_w $
induced on $ \esh \subset \bsh $ by the weak operator topology is the
coarsest topology on $ \esh $ such that the restrictions of the linear
functionals (\ref{afun}) to $ \esh $ are continuous. Since these restrictions
are given by
\[
P \mapsto \ip{\fii}{P\fii} = \tr{PQ} = h_Q(P)
\]
where $ P \in \esh $ and $ Q := \kb{\fii}{\fii} \in \esh $, the topology
$ \mT_w $ on $ \esh $ is, according to (\ref{p}), just our topology $ \mT_0 $.

Now we compare the weak topology $ \mT_0 $ with the metric topology
$ \mT_n $.

\begin{theorem}\label{thm:top-esh}
The weak topology $ \mT_0 $ on $ \esh $ and the metric topology $ \mT_n $ on
$ \esh $ are equal.
\end{theorem}
\proof{
According to (\ref{p}), a neighborhood base of $ P \in \esh $ w.r.t.\ $ \mT_0 $
is given by the open sets
\begin{equation}\label{up}
\begin{array}{c}
U(P;Q_1,\ldots,Q_n;\varepsilon) \hspace{7.5cm}                \vspace{2mm}\\
\hspace{0.8cm}
\begin{array}{crl}
& := & \displaystyle{\bigcap_{i=1}^n h_{Q_i}^{-1}
        ( \, ]h_{Q_i}(P) - \varepsilon,h_{Q_i}(P) + \varepsilon[ \, )}
                                                              \vspace{2mm}\\
&  = & \bigl\{ \widetilde{P} \in \esh \, \bigl| \, \bigl|
        h_{Q_i}(\widetilde{P}) - h_{Q_i}(P) \bigr| < \varepsilon \
        {\rm for} \ i=1,\ldots,n \bigr\}                      \vspace{2mm}\\
&  = & \bigl\{ \widetilde{P} \in \esh \, \bigl| \, \bigl|
        \tr{\widetilde{P}Q_i} - \tr{PQ_i} \bigr| < \varepsilon   \
        {\rm for} \ i=1,\ldots,n \bigr\}
\end{array}
\end{array}
\end{equation}
where $ Q_1,\dots,Q_n \in \esh $ and $ \varepsilon > 0 $;
a neighborhood base of $P$ w.r.t.\ $ \mT_n $ is given by the open balls
\begin{equation}\label{kep}
K_{\varepsilon}(P) := \bigl\{ \widetilde{P} \in \esh \, \bigl| \,
                      \bigl\| \widetilde{P} - P \bigr\| < \varepsilon
                                                        \bigr\}.
\end{equation}
If $ \bigl\| \widetilde{P} - P \bigl\| < \varepsilon $, then
\[
\bigl| \tr{\widetilde{P}Q_i} - \tr{PQ_i} \bigr|
  =   \bigl| \tr{Q_i(\widetilde{P} - P)} \bigr|
 \leq \no{Q_i}_{\mathrm{tr}} \bigl\| \widetilde{P} - P \bigr\|
  =   \bigl\| \widetilde{P} - P \bigr\|
  <   \varepsilon;
\]
hence, $ K_{\varepsilon}(P) \subseteq U(P;Q_1,\ldots,Q_n;\varepsilon) $. To
show some converse inclusion, take account of Theorem \ref{thm:esh-dist}, part (a), and note
that
\[
\bigl\| \widetilde{P} - P \bigr\|^2
  = 1 - \tr{\widetilde{P}P}
  = \bigl| \tr{\widetilde{P}P} - \tr{PP} \bigr|.
\]
In consequence, by (\ref{up}) and (\ref{kep}),
$ U(P;P;\varepsilon^2) = K_{\varepsilon}(P) $. Hence,
$ \mT_0 = \mT_n $. \qed}

It looks surprising that the topolgies $ \mT_0 $ and
$ \mT_n $ coincide. In fact, consider the sequence
$ \{ P_{\fii_n} \}_{n \in \N} $ where the vectors
$ \fii_n \in \hi $ constitute an orthonormal system. Then,
w.r.t.\ the weak operator topology, $ P_{\fii_n} \to 0 $ as
$ n \to \infty $ whereas $ \no{P_{\fii_n} - P_{\fii_{n+1}}} =1 $
for all $ n \in \N $. However, $ 0 \not\in \esh $; so
$ \{ P_{\fii_n} \}_{n \in \N} $ is convergent neither w.r.t.\
$ \mT_w = \mT_0 $ nor w.r.t.\ $ \mT_n $. Finally, like in the case
of the weak operator topology, there is a natural uniform structure
inducing $ \mT_0 $. The uniform structures that are canonically related to
$ \mT_0 $ and $ \mT_n $ are different: $ \{ P_{\fii_n} \}_{n \in \N} $ is
a Cauchy sequence w.r.t.\ the uniform structure belonging to $ \mT_0 $
but not w.r.t.\ that belonging to $ \mT_n $, i.e., w.r.t.\ the metric
$ \rho_n $.

We remark that besides $ \mT_0 $ and $ \mT_w $ several further
weak topologies can be defined on $ \esh $. Let $ \csh $ be the Banach space of the compact
self-adjoint operators and remember that $ (\csh)' = \tsh $. So the weak
Banach-space topologies of $ \csh $, $ \tsh $, and $ \bsh $ as well as
the weak-* Banach-space topologies of $ \tsh $ and $ \bsh $ can
be restricted to $ \esh $, thus giving the topologies
$ \mT_1 := \sigma(\csh,\tsh) \cap \esh $,
$ \mT_2 := \sigma(\tsh,\csh) \cap \esh $,
$ \mT_3 := \sigma(\tsh,\bsh) \cap \esh $,
$ \mT_4 := \sigma(\bsh,\tsh) \cap \esh $, and
$ \mT_5 := \sigma(\bsh,(\bsh)') \cap \esh $. Moreover,
the strong operator topology induces a topology $ \mT_s $ on
$ \esh $. From the obvious inclusions
\[
\mT_w \subseteq \mT_1 \subseteq \mT_2 \subseteq \mT_3
      \subseteq \mT_{\mathrm{tr}} ,
\]
\[
\mT_1 = \mT_4 \subseteq \mT_5 = \mT_1 ,
\]
and
\[
\mT_w \subseteq \mT_s \subseteq \mT_n
\]
as well as from the shown equality
\[
\mT_0 = \mT_w = \mT_n = \mT_{\mathrm{tr}}
\]
it follows that the topologies $ \mT_1,\ldots,\mT_5 $ and $ \mT_s $
also coincide with $ \mT_0 $.

Finally, we show that all the topologies on $ \esh $ are equivalent to
the quotient topologies $ \mT_{\mu} $ and $ \mT_{\nu} $ on $ \ph $, resp.,
$ S/S^1 $.

\begin{theorem}\label{thm:ss1-esh}
The mapping $ F \! : S/S^1 \to \esh $, $ F([\fii]_S := P_{\fii} $ where
$ \fii \in S $, is a homeomorphism between the topological spaces
$ (S/S^1,\mT_{\nu}) $ and $ (\esh,\mT_0) $.
\end{theorem}
\proof{
The mapping $F$ is bijective. The map
$ h_Q \circ F \circ \nu \! : S \to \R $
where $ h_Q $ is any of the functions given by Eq.~(\ref{p}) and
$ \nu $ is the canonical projection from $S$ onto $ S/S^1 $,
reads explicitly
\[
(h_Q \circ F \circ \nu)(\fii) = h_Q(F([\fii]_S)) = h_Q(P_{\fii})
                              = \tr{P_{\fii}Q}   = \ip{\fii}{Q\fii};
\]
therefore, $ h_Q \circ F \circ \nu $ is continuous. Consequently,
for an open set $ O \subseteq \R $,
\[
(h_Q \circ F \circ \nu)^{-1}(O) = \nu^{-1}(F^{-1}(h_Q^{-1}(O)))
\]
is an open set of $S$. By the definition of the quotient topology
$ \mT_{\nu} $, it follows that $ F^{-1}(h_Q^{-1}(O)) $ is an open set of
$ S/S^1 $. Since the sets $ h_Q^{-1}(O) $, $ Q \in \esh $,
$ O \subseteq \R $ open, generate the weak topology $ \mT_0 $,
$ F^{-1}(U) $ is open for any open set $ U \in \mT_0 $. Hence,
$F$ is continuous.

To show that $F$ is an open mapping, let $ V \in \mT_{\nu} $ be an open
subset of $ S/S^1 $ and let $ [\fii_0]_S \in V $. Since the canonical
projection $ \nu $ is continuous, there exists an $ \varepsilon > 0 $
such that
\begin{equation}\label{nuke}
\nu(K_{\varepsilon}(\fii_0) \cap S) \subseteq V
\end{equation}
where $ K_{\varepsilon}(\fii_0) := \{ \fii \in \hi \, | \,
\no{\fii - \fii_0} < \varepsilon \} $. Without loss of generality
we assume that $ \varepsilon < 1 $.

The topology $ \mT_0 $ is generated by the functions $ h_Q $
according to (\ref{p}); $ \mT_0 $ is also generated by the functions
$ P \mapsto g_Q(P) := \sqrt{h_Q(P)} = \sqrt{\tr{PQ}} $. In consequence,
the set
\[
U_{\varepsilon}
 :=  g_Q^{-1} \left( \, \left] 1 - \tfrac{\varepsilon}{2},
                               1 + \tfrac{\varepsilon}{2} \right[ \, \right)
\cap h_Q^{-1} \left( \, \left] 1 - \tfrac{\varepsilon^2}{4},
                               1 + \tfrac{\varepsilon^2}{4} \right[
                                                                  \, \right)
\]
where $ Q := P_{\fii_0} $ and $ \fii_0 $ and $ \varepsilon $ are specified
in the preceding paragraph, is $ \mT_0 $-open. Using the identity
\[
1 - |\ip{\fii_0}{\fii}|^2 = \no{\fii - \ip{\fii_0}{\fii} \fii_0}^2
\]
where $ \fii \in \hi $ is also a unit vector, we obtain
\begin{eqnarray*}
U_{\varepsilon}
& = & \left\{ P_{\fii} \in \esh \left| \,
        |g_Q(P_{\fii}) - 1| < \tfrac{\varepsilon}{2} \ {\rm and} \
        |h_Q(P_{\fii}) - 1| < \tfrac{\varepsilon^2}{4} \right. \right\}   \\
& = & \left\{ P_{\fii} \in \esh \left| \,
      \bigl| |\ip{\fii_0}{\fii}| - 1 \bigr| < \tfrac{\varepsilon}{2} \
                                                           {\rm and} \
      \bigl| |\ip{\fii_0}{\fii}|^2 - 1 \bigr| < \tfrac{\varepsilon^2}{4}
                                                       \right. \right\}   \\
& = & \left\{ P_{\fii} \in \esh \left| \,
      \bigl| |\ip{\fii_0}{\fii}| - 1 \bigr| < \tfrac{\varepsilon}{2} \
                                                           {\rm and} \,
      \no{\fii - \ip{\fii_0}{\fii} \fii_0} < \tfrac{\varepsilon}{2}
                                                       \right. \right\}.
\end{eqnarray*}
Now let $ P_{\fii} \in U_{\varepsilon} $. Since $ \varepsilon < 1 $, we have
that $ \ip{\fii}{\fii_0} \neq 0 $. Defining the phase factor
$ \lambda := \frac{\ip{\fii}{\fii_0}}{|\ip{\fii}{\fii_0}|} $,
it follows that
\begin{eqnarray*}
\no{\lambda \fii - \fii_0}
& = & \no{\lambda \fii - \lambda \ip{\fii_0}{\fii} \fii_0}
        + \no{\lambda \ip{\fii_0}{\fii} \fii_0 - \fii_0}         \\
& = & \no{\fii - \ip{\fii_0}{\fii} \fii_0}
        + \bigl\| |\ip{\fii_0}{\fii}| \fii_0 - \fii_0 \bigr\|    \\
& < & \tfrac{\varepsilon}{2} + \tfrac{\varepsilon}{2}            \\
& = & \varepsilon.
\end{eqnarray*}
That is, $ P_{\fii} \in U_{\varepsilon} $ implies that
$ \lambda \fii \in K_{\varepsilon}(\fii_0) $; moreover,
$ \lambda \fii \in K_{\varepsilon}(\fii_0) \cap S $.

Taking the result (\ref{nuke}) into account, we conclude that,
for $ P_{\fii} \in U_{\varepsilon} $,
$ [\fii]_S = [\lambda \fii]_S = \nu(\lambda \fii) \in V $. Consequently,
$ P_{\fii} = F([\fii]_S) \in F(V) $. Hence,
$ U_{\varepsilon} \subseteq F(V) $. Since $ U_{\varepsilon} $
is an open neighborhood of $ P_{\fii_0} $, $ P_{\fii_0} $
is an interior point of $ F(V) $. So, for every $ [\fii_0]_S \in V $,
$ F([\fii_0]_S) = P_{\fii_0} $ is an interior point of $ F(V) $, and
$ F(V) $ is a $ \mT_0 $-open set. Hence, the continuous bijective map
$F$ is open and thus a homeomorphism. \qed}

In the following, we identify the sets $ \ph $, $ S/S^1 $, and $ \esh $
and call the identified set the {\it projective Hilbert space
$ \ph $}. However, we preferably think about the elements of
$ \ph $ as the one-dimensional orthogonal projections $ P = P_{\fii} $. On
$ \ph $ then the quotient topologies $ \mT_{\mu} $, $ \mT_{\nu} $, the
weak topologies $ \mT_0 $, $ \mT_w $, $ \mT_1,\ldots,\mT_5 $, $ \mT_s $,
and the metric topologies $ \mT_n $, $ \mT_{\mathrm{tr}} $ coincide. So
we can say that $ \ph $ carries a natural topology $ \mT $;
$ (\ph,\mT) $ is a second-countable Hausdorff space.

For our purposes, it is suitable to represent this topology $ \mT $ as
$ \mT_0 $, $ \mT_n $, or $ \mT_{\mathrm{tr}} $. As already discussed,
the topologies $ \mT_0 $, $ \mT_n $, and $ \mT_{\mathrm{tr}} $ are
canonically related to uniform structures. With respect to the uniform
structure inducing $ \mT_0 $, $ \ph $ is not complete. The uniform
structures related to $ \mT_n $ and $ \mT_{\mathrm{tr}} $ are the same
since they are induced by the equivalent metrics $ \rho_n $ and
$ \rho_{\mathrm{tr}} $; $ (\ph,\rho_n) $ and $ (\ph,\rho_{\mathrm{tr}}) $
are separable complete metric spaces. So $ \mT $ can be defined by
a complete separable metric, i.e., $ (\ph,\mT) $ is a polish space.

\section{The Measurable Structure of $\ph$}\label{sec:meas}

It is almost natural to define a measurable structure on the
projective Hilbert space $ \ph $ by the $ \sigma $-algebra $ \Xi =
\Xi(\mT) $ generated by the $ \mT $-open sets, i.e., $ \Xi $ is the
smallest $ \sigma $-algebra containing the open sets of the natural
topology $ \mT $. In this way $ (\ph,\Xi) $ becomes a measurable
space where the elements $ B \in \Xi $ are the Borel sets of $ \ph
$. However, since the topology $ \mT $ is generated by the
transition-probability functions $ h_Q $ according to Eq.\
(\ref{p}), it is also obvious to define the measurable structure of
$ \ph $ by the $ \sigma $-algebra $ \Sigma $ generated by the
functions $ h_Q $, i.e., $ \Sigma $ is the smallest $ \sigma
$-algebra such that all the functions $ h_Q $ are measurable. A
result due to Misra (1974) \cite[Lemma 3]{mis74} clarifies the
relation between $ \Xi $ and $ \Sigma $. Before stating that result,
we recall the following simple lemma which we shall also use later.

\begin{lemma}\label{lem:sigt-sigb}
Let $ (M,\mT) $ be any second-countable topological space,
$ \mB \subseteq \mT $ a countable base, and $ \Xi = \Xi(\mT) $ the
$ \sigma $-algebra of the Borel sets of $M$. Then
$ \Xi = \Xi(\mT) = \Xi(\mB) $ where $ \Xi(\mB) $ is the
$ \sigma $-algebra generated by $ \mB $; $ \mB $ is
a countable generator of $ \Xi $.
\end{lemma}
\proof{
Clearly, $ \Xi(\mB) \subseteq \Xi(\mT) $. Since every open set
$ U \in \mT $ is the countable union of sets of $ \mB $, it follows that
$ U \in \Xi(\mB) $. Therefore, $ \mT \subseteq \Xi(\mB) $ and consequently
$ \Xi(\mT) = \Xi(\mB) $. \qed}

\begin{theorem}[{\rm Misra}]\label{thm:misra}
The $ \sigma $-algebra $ \Xi = \Xi(\mT) $ of the Borel sets of the
projective Hilbert space $ \ph $ and the $ \sigma $-algbra $ \Sigma $
generated by the transition-probability functions $ h_Q $, $ Q \in \ph $,
are equal.
\end{theorem}
\proof{
Since $ \mT $ is generated by the functions $ h_Q $, the latter are
continuous and consequently $ \Xi $-measurable. Since $ \Sigma $ is
the smallest $ \sigma $-algebra such that the functions $ h_Q $ are
measurable, it follows that $ \Sigma \subseteq \Xi $.

Now, by Lemma \ref{lem:esh}, $ \mT $ is second-countable, and a countable base
$ \mB $ of $ \mT $ is given by the finite intersections of the sets
$ U_{klm} $ according to Eq.\ (\ref{uklm}). Since $ U_{klm} \in \Sigma $,
it follows that $ \mB \subseteq \Sigma $. By Lemma \ref{lem:sigt-sigb}, we conclude that
$ \Xi = \Xi(\mB) \subseteq \Sigma $. Hence, $ \Xi = \Sigma $. \qed}

We remark that our proof of Misra's theorem is much easier than
Misra's proof from 1974. The reason is that we explicitly used the
countable base $ \mB $ of $ \mT $ consisting of $ \Sigma $-measurable sets.

Finally, consider the $ \sigma $-algebra $ \Xi_0 $ in $ \ph $ that is
generated by all $ \mT $-continuous real-valued functions on $ \ph $, i.e.,
$ \Xi_0 $ is the $ \sigma $-algebra of the Baire sets of $ \ph $. Obviously,
$ \Sigma \subseteq \Xi_0 \subseteq \Xi $; so Theorem \ref{thm:misra} implies that
$ \Xi_0 = \Xi $. This result is, according to a general theorem,
also a consequence of the fact that the topology $ \mT $ of
$ \ph $ is metrizable.

Summarizing, our result $ \Sigma = \Xi_0 = \Xi $ manifests that
the projective Hilbert space carries, besides its natural topology
$ \mT $, also a very natural measurable structure $ \Xi $.


\begin{thebibliography}{99}
\bibitem{bel95a;b}
Beltrametti, E. G., and S. Bugajski, ``A Classical Extension of
Quantum Mechanics,'' {\it J. Phys.\ A:\ Math.\ Gen.}\ {\bf 28},
3329--3343 (1995); ``Quantum Observables in Classical Frameworks,''
{\it Int.\ J. Theor.\ Phys.}\ {\bf 34},
1221--1229 (1995).

\bibitem{bje05}
Bjelakovi\'c, I., and W. Stulpe, ``The Projective Hilbert Space as
a Classical Phase Space for Nonrelativistic Quantum Dynamics,''
{\it Int.\ J. Theor.\ Phys.}\ {\bf 44}, 2041--2049 (2005).

\bibitem{bro01} Brody, D. C., and L. P. Hughston, ``Geometric
Quantum  Mechanics," {\it J. Geom.\ Phys.}\ {\bf 38}, 19--53 (2001).

\bibitem{bug91;93a-d}
Bugajski, S., ``Nonlinear Quantum Mechanics is a Classical Theory,''
{\it Int.\ J. Theor.\ Phys.}\ {\bf 30}, 961--971 (1991);
``Delinearization of Quantum Logic,''
{\it Int.\ J. Theor.\ Phys.}\ {\bf 32}, 389--398 (1993);
``Classical Frames for a Quantum Theory---A Bird's-Eye View,''
{\it Int.\ J. Theor.\ Phys.}\ {\bf 32}, 969--977 (1993);
``On Classical Representations of Convex Descriptions,''
{\it Z. Naturforsch.}\ {\bf 48a}, 469--470 (1993).

\bibitem{bug94}
Bugajski, S., ``Topologies on Pure Quantum States,''
{\it Phys.\ Lett.\ A}\ {\bf 190}, 5--8 (1994).

\bibitem{bus04}
Busch, P., ``Less (Precision) Is More (Information): Quantum Information
in Terms of Quantum Statistical Models,'' quant-ph/0401027 (2004).

\bibitem{bus07}
Busch, P., and W. Stulpe, ``The Structure of Classical Extensions
of Quantum Probability Theory,'' Preprint (2007).

\bibitem{cir84}
Cirelli, R., and P. Lanzavecchia, ``Hamiltonian Vector Fields in
Quantum Mechanics,'' {\it Nuovo Cim.}\ {\bf 79 B}, 271--283 (1984).

\bibitem{cir90a;b}
Cirelli, R., A. Mania, and L. Pizzocchero, ``Quantum Mechanics as an
Infinite-Dimensional Hamiltonian System with Uncertainty
Structure,'' Parts I and II, {\it J. Math.\ Phys.}\ {\bf 31},
2891--2897, 2898--2903 (1990).

\bibitem{gue77}
G\"unther, C., ``Prequantum Bundles and Projective Hilbert
Geometries," {\it Int.\ J. Theor.\ Phys.}\ {\bf 16}, 447--464
(1977).

\bibitem{hol82}
Holevo, A. S., {\it Probabilistic and Statistical Aspects of
Quantum Theory}, North Holland, Amsterdam (1982).

\bibitem{kib79}
Kibble, T. W. B., ``Geometrization of Quantum Mechanics,"
{\it Commun.\ Math.\ Phys.}\ {\bf 65}, 189--201 (1979).

\bibitem{mis74}
Misra, B., ``On a New Definition of Quantal States,'' in {\it Physical
Reality and Mathematical Description}, C. P. Enz and J. Mehra (eds.),
455--476, Reidel, Dordrecht (1974).

\bibitem{stu01}
Stulpe, W., and M. Swat, ``Quantum States as Probability Measures,''
{\it Found.\ Phys.\ Lett.}\ {\bf 14}, 285--293 (2001).

\end{thebibliography}
\end{document}